\documentclass[10pt]{article}

\usepackage{amssymb} 
\usepackage{comment} 
\usepackage{color} 
\usepackage{graphicx} 
\usepackage{rotating} 

\def\today{Theoretical and mathematical physics, accepted 07 April 2025} 

\def \PVI    {{\rm P_{\rm VI}}}
\def \Pn     {{\rm P_{\rm n}}}

\def \fold    {<=}
\def \fold    {,}

\def \CRAS{C.~R.~Acad.~Sc.~Paris}
\def \ccomma{\raise 2pt\hbox{,}\ } 
\def \D {\hbox{d}}

\def\vect{\mbox{\boldmath{$\theta$}}}
\def\vecT{\mbox{\boldmath{$\Theta$}}}

\begin{document}


\title{Minimal algebraic solutions of the sixth equation of Painlev\'e}

\author{Robert Conte${}^{1,2}$
{}\\
\\ 1. Universit\'e Paris-Saclay, ENS Paris-Saclay, CNRS,
\\ Centre Borelli, F-91190 Gif-sur-Yvette, France.
\\
\\ 2. Department of Mathematics, The University of Hong Kong,
\\ Pokfulam, Hong Kong.
}

\maketitle



AMS MSC 2000 
33E17 

\begin{abstract}
For each of the forty-eight exceptional algebraic solutions $u(x)$ of the sixth equation of Painlev\'e,
we build the algebraic curve $P(u,x)=0$ of a degree conjectured to be minimal,
then we give an optimal parametric representation of it.
This degree is equal to the number of branches, except for fifteen solutions.
\end{abstract}

\tableofcontents

\section{Introduction and motivation}

Apart the transcendental general solution which characterizes it,
the sixth equation $\PVI$ of Painlev\'e
\begin{eqnarray}
& & {\hskip -20.0 truemm}
\frac{\D^2 u}{\D x^2}=
 \frac{1}{2} \left[\frac{1}{u} + \frac{1}{u-1} + \frac{1}{u-x} \right] \left(\frac{\D u}{\D x}\right)^2
- \left[\frac{1}{x} + \frac{1}{x-1} + \frac{1}{u-x} \right] \frac{\D u}{\D x}
\nonumber \\ & & {\hskip -20.0 truemm} \phantom{123456}
+ \frac{u (u-1) (u-x)}{2 x^2 (x-1)^2}
  \left[\theta_\infty^2 - \theta_0^2 \frac{x}{u^2} + \theta_1^2 \frac{x-1}{(u-1)^2}
        + (1-\theta_x^2) \frac{x (x-1)}{(u-x)^2} \right],
\nonumber \\ & & {\hskip -20.0 truemm}
(2\alpha, -2\beta, 2\gamma, 1-2\delta)=(\theta_\infty^2,\theta_0^2,\theta_1^2,\theta_x^2),
\label{eqPVI}
\end{eqnarray} 
admits exactly 48 exceptional algebraic solutions $u(x)$,
in the sense that they are
neither those mentioned by Picard 
\cite[p.~298]{Picard1889} when the four constant parameters $\theta_j$ vanish,
nor some particular solutions of the Riccati equation admitting $\PVI$ as
a differential consequence for some constraint between the $\theta_j$'s.
They were discovered by various authors~:
Hitchin \cite{Hitchin1995-Poncelet}, 
Dubrovin \cite{DubLNM}, 
Dubrovin and Mazzocco \cite{DM2000}, 
Kitaev \cite{Kitaev2005Dessins} 
       \cite{Kitaev2006Angers}, 
Andreev and Kitaev \cite{Andreev-Kitaev-PVI-2002}, 
and all the others by
Boalch 
\cite{Boalch-Icosa}
\cite{Boalch2007Bolibrukh}
\cite{Boalch2005Klein}
\cite{Boalch2007Highergenus}.
The proof that no other such solutions exist is due to 
Lisovyy and Tykhyy \cite{LT2014}.

\medskip

These 48 solutions 
are in fact equivalence classes of a group which leaves $\PVI$ form-invariant
without constraining the $\theta_j$'s, 
group made of:
the 4 sign changes of the $\theta_j$'s, 
the 4! homographies of $(u,x)$ 
and a unique birational transformation  (see Appendix).
Let us denote $P(u,x)=0$ the algebraic curve representing an element of one of these equivalence classes.
Its genus $g$ and number of branches $b$ (degree of the polynomial $P$ in $u$) are invariant under the group,
but its degree $d$ (global degree of $P$ in $u$ and $x$) is not.
In order to simplify the writing of the solutions,
it is therefore convenient to lower this degree $d$, ideally to its minimal value $b$,
by the repeated action of elements of the group.
\medskip

\textit{Remark}.
The above notation $\vect=(\theta_\infty,\theta_0,\theta_1,\theta_x)$
for the sequence of the four monodromy exponents
is the natural choice for at least three reasons.
The first one is the writing of $\PVI$ in elliptic coordinates\cite{FuchsP6,PaiCRAS1906,CMBook2}
\begin{eqnarray}
& &
\frac{\D^2 U}{\D X^2}=\frac{(2 \omega)^{3}}{\pi^2}
\sum_{j=\infty,0,1,x}\theta_j^2 \wp'(2\omega U+\omega_j,g_2,g_3),
\label{eqPVIUX}
\end{eqnarray}
obviously form-invariant under any permutation of the four half-periods
$\omega_j$ of the elliptic function $\wp$;
defining $\theta_\infty$ with a shift of unity by $2\alpha=(\theta_\infty-1)^2$ would break this invariance.
The second one is the classical confluence \cite{PaiCRAS1906} between the $\Pn$'s,
in which the four singularities $(\infty,0,1,x)$ of $u$ 
become successively $(\infty,0,1)$, $(\infty,0)$, $(\infty)$,
which defines an order in the sequence $(\theta_\infty,\theta_0,\theta_1,\theta_x)$;
reordering this sequence 
into $(\theta_0,\theta_x,\theta_1,\theta_\infty-1)$ \cite[Eq.~(4)]{Boalch-Icosa} or
into $(\theta_0,\theta_1,\theta_x,\theta_\infty-1)$ \cite[p.~125]{LT2014} 
would make one forget about this natural ordering.
The third one is the choice of Painlev\'e \cite{PaiCRAS1906}
for the Greek alphabetical order $\alpha,\beta,\gamma,\delta$.

\section{Statement of the problem}

The goal of the present paper is thefore to solve the following problem.

\textbf{Problem}. For each of the $48$ algebraic solutions,
find a representative in its equivalence class whose degree $d$ equals the number of branches $b$.

As an example, consider one of the two solutions with $b=5$ branches, 
the one found by Kitaev \cite[Eq.~(3.3)]{Kitaev2005Dessins}
which bears the number I21 in the tables de Boalch \cite{Boalch-Icosa} and 2 in those of Lisovyy and Tykhyy \cite{LT2014}.
Its genus is $g=0$, its degree the nonoptimal value $d=b+1$ and its number of terms is 12.
An homography followed by the birational transformation 
simplifies it to $b=d=5$ containing only 4 terms,
\begin{eqnarray}
& &
\left\lbrace
\begin{array}{ll}
    \displaystyle{\vect=\frac{(3,1,2,1)}{5}, P=9 (2-x) u^5 + 15 (x -4) u^4 +5 (10-x^2+9 x) u^3 -15 (5 x+3 x^2) u^2	
}\\ \displaystyle{\phantom{12345678901234567890}		+45 x^2 u-x^2-8 x^3, 
}\\ \displaystyle{\vect=\frac{(1,1,2,3)}{5}, P=(9 x-8) u^5-5 u^4 x-5 x (x-3) u^3 - 5 x (3 x-1) u^2+5 x^2 u-9 x^2+8 x^3,
}\\ \displaystyle{\vect=\frac{(0,0,1,2)}{5}, P=u^5           -5 x      u^3+5 x   u^2-x^2.
}
\end{array}
\right.
\end{eqnarray} 

The sought simplification therefore consists in removing any fixed pole of $u$
(in the above example I21, the fixed pole $x=2$),
so that $u$ admits as only singularities the three critical fixed points $x=\infty,0,1$.
For such removable fixed singularities,
see for example the figures in \cite{PCI}.

There exists another transformation \cite{Kitaev1991}, called folding transformation, 
and it is unique \cite{TOS},
which leaves $\PVI$ form-invariant,
but this is at the expense of two constraints between the $\theta_j$'s.
A solution $\vect=(0,0,2 a, 2 b)$ (\textit{modulo} permutations and sign changes)
is folded into another solution $\vect=(a,a,b,b)$,
an operation symbolized as ``unfolded $\le$ folded'' par Boalch.


Let us recall the partition of the 48 algebraic solutions 
in three disjoint types \cite[p.~233]{Boalch2010}~:
\begin{enumerate}
	\item (genus zero) three solutions which depend on at most two arbitrary $\theta_j$'s;
	\item (genus zero or one) thirty solutions
having rational, non-arbitrary $\theta_j$'s,
inequivalent under the folding transformation \cite{Kitaev1991};
	\item (genera 0, 1, 2, 3, 7) fifteen transformed of seven out of the thirty ones by folding(s).
\end{enumerate}

The final result is the following. 

\textbf{Conjecture}. 
Only 
$7$ of the $30$ unfolded solutions, and
$8$ of the $15$ folded ones
display a nonzero difference $d-b$,
and this difference is minimal. 

Before the present work, only
$4$ of the $30$ unfolded solutions, and
$3$ of the $15$ folded ones
displayed a minimal $d-b$.
The main reason is precisely the elegance of the method of Boalch.
That method makes him find 
rational numbers $\theta_j$ whose denominators have a minimal gcd \cite[Table 2]{Boalch-Icosa},
and it happens that the birational transformation doubles this gcd in most cases
but that, surprisingly, it greatly simplifies (in the sense above defined) the representative.
Table \ref{TableNotation} displays numerous such examples~: I22, I23, O09, I24, I25, etc.

Several facts support this conjecture.

\begin{enumerate}
	
	\item 
	Existence of several sets of siblings 
(in the sense defined by Boalch \cite{Boalch-Icosa}, 
i.e.~whose all members have by definition the same values of $b$ and $x$),
one member having an optimal degree $d=b$ and the others degrees $d \ge b+1$ 
impossible to lower by any transformation.
	
	\item 
Probable nonexistence,
for the solution I50 with $b=40$ branches,
of a representative $d-b \le 5$ whose all $\theta_j$'s would be equal,
so as to preserve its property to be doubly folded.
Indeed,
the representatives whose common value of the four $\theta_j$'s
is respectively $3/20,7/20,13/20,17/20$
have degrees $46, 54, 78, 102$ 
and a number of terms equal to $335, 663, 1647, 2631$.

	\item 
Failure to find a representative $d=b$
of the solution with the smallest $b$ having a degree $d \not=b$ (solution labeled K= Klein),
after a triple loop 
on the $2^4$ sign changes, the $4!$ homographies
and five birational transformations 
(three months of computation).

\end{enumerate}

\vfill\eject

\section{Results}

They take two forms.

\begin{enumerate}
	\item 
	The solution of the above stated problem,
summarized in Table \ref{TableNotation}~:
among the $30+15$ solutions,
the $7+8$ without a representative $d=b$ have a minimal number $d-b$ of fixed poles equal to 
six for I50, I52,
two for I38, 238, I46, O13, I51, I47,
one for K, I34, I37, I43, I28, O12, I48.
In particular, the associated representative possesses the maximal possible number of null $\theta_j$'s.
	
	\item 
	For a certain representative of each solution,
a ``simple'' (in the senses of Appendix \ref{sectionSimple}) parametric representation  
listed in sections \ref{sectiong0} and \ref{sectiong1}.

\end{enumerate}

The representatives of these two forms may not be identical,
but they only differ by a homography.

Table \ref{TableNotation} deserves some remarks.
\begin{enumerate}

	\item 
	All minimal solutions have all their $\theta_j$'s smaller than the unity.
		
	\item 
	The strategy of Dubrovin and Mazzocco \cite{DM2000}
consisting in looking for solutions with three null $\theta_j$,
which led them to the discovery of solutions (H3), (H3)', (H3)''
(Table \ref{TableNotation}, last column),
perfectly matches the property to be doubly foldable,
a property which requires the presence of at least two null $\theta_j$'s.

	\item 
	Only four solutions, all doubly folded (O13, I50, I51, I52),
are invariant under the 24 homographies, their $\theta_j$'s are all equal.
	
	\item 
  Only one solution (I33) is not invariant by any homography (apart the identity),
this is the one called ``generic'' by Boalch. 
This is also the unique one whose four $\theta_j$'s are all different,
whatever be the representative.
	
	\item 
	As noticed by Boalch \cite{Boalch2007Highergenus},
in each set of siblings
(I22, I23), (I24, I25), (I26, I27), (I29, I30), (I34, I35), (I37, I38), (I39, I40), (237, 238, 239), (I42, I43), (I44, I45), (I47, I48), (I50,I51),
all the members
can be represented by the same value of $x$,
at the possible expense of a greater degree or a higher number of terms.
For instance, the minimal number of terms in the equivalence class of I24 is 11, not 17. 

	\item 
	The siblings (I26, I27) ($b=d=9$) seem to have not been noticed yet.
	
	\item  
	The solution I36 and the siblings (I42, I43) share the same elliptic curve,
a fact yet unnoticed.
	
	\item 
  Two quadruplets of monodromy exponents, 
respectively $(0,0,1,1)/6$ and $(1,1,1,1)/12$,
are common to more than one solution,
respectively (T06, O11, I49) and (O12, I52).

	\item 
  The solution III has only one child, but this is the parent of O11 and T06.
	
	\item 
  All solutions with two null $\theta_j$'s are minimal.
\end{enumerate}

\baselineskip=12truept

\hfill

{\vglue 20.0 truemm}

\tabcolsep=1.5truemm
\tabcolsep=0.5truemm

\begin{sidewaystable}[ht]
\caption[The 3+30+15 canonical exceptional solutions]
{        The 3+30+15 minimal   exceptional solutions. \hfill \today \\
Columns~: notation of Refs \cite{Boalch-Icosa} and \cite{LT2014},
genus $g$ (preceded by H (hyperelliptic) or N (non-hyperelliptic) if $g>1$) of the curve $P(u,x)=0$,
number $d-b$ of fixed poles if nonzero, 
number of terms of $P$,
sequence $\vect=(\theta_j)$,
chain of foldings (filiation),
set of siblings,
list of homographies leaving invariant the curve $P=0$
(their number is a divisor of 24 except 12, the identity is omitted),
former best representative ($d-b$, number of terms, $(\theta_j)$),
comment.
}
 \vspace{0.0truecm}
\begin{center}
{\footnotesize
\begin{equation}
\begin{array}{| c | c | r | r | r | r | l | l | l | l || l | l |}
\hline 
\hbox{B}&\hbox{LT}&g&b&d-b&\hbox{terms}& (\theta_\infty,\theta_0,\theta_1,\theta_x) & \hbox{filiation}& \hbox{siblings}
        & \hbox{homographies} & \hbox{former }d-b, \hbox{ terms, }\theta_j & \hbox{comment} \\ \hline
\hline
   &II & 0 & 2 &    &   2 &(  a,  a,b,b) & II \fold II          &               & 2,7,8,15,16,21,22 &              & \hbox{}  \\ \hline 
   &III& 0 & 3 &    &   3 &(a,2 a,a,1/3) & III\fold T06 \hbox{ }&               &18            &                   & \hbox{ cube}  \\ \hline  
   &   &   &   &    &     &              & III\fold O11\fold O12&               &              &                   & \hbox{ cube}  \\ \hline 
   &IV & 0 & 4 &    &   4 &(b,b,b,1/2)   & IV \fold O10\fold O13&               &3,8,12,14,18  &                   & \hbox{ tetrahedron}  \\ \hline 
\hline
I20&  1& 0 & 5 &    &   7 &(1,3,3,4)/15  &                      &               & 3            & 1,13,(5,6,3,5)/15 & \hbox{}  \\ \hline  
I21&  2& 0 & 5 &    &   4 &(0,0,1,2)/5   & I21\fold I28         &               & 8            & 1,12,(3,1,2,1)/5  & \hbox{}  \\ \hline 
O08&  5& 0 & 6 &    &   5 &(0,0,1,5)/12  & O08\fold O07         &               & 8            & 0,14,(3,4,4,9)/12 & \hbox{}  \\ \hline 
I23&  6& 0 & 6 &    &   8 &(1,5,7,7)/30  &                      & I22, I23      & 2            & 2,16,(5,6,3,6)/15 & \hbox{}  \\ \hline       
I22&  7& 0 & 6 &    &   7 &(1,1,5,13)/30 &                      & I22, I23      & 8            & 4,23,(10,3,6,3)/15& \hbox{}  \\ \hline    
 K &  8& 0 & 7 &  1 &  12 &(1,1,2,1)/7   &                      &               & 6,8,10,19,23 & 3,24,(3,2,2,2)/7  & \hbox{}  \\ \hline 
O09& 10& 0 & 8 &    &  13 &(1,3,3,7)/24  &                      &               & 3            & 2,24,(4,4,6,3)/12 & \hbox{}  \\ \hline    
I24& 11& 0 & 8 &    &  11 &(1,1,7,5)/20  &                      & I24, I25      & 8            & 1,30,(2,5,2,4)/10 & \hbox{} 	\\ \hline 
I25& 12& 0 & 8 &    &  12 &(1,3,5,3)/20  &                      & I24, I25      & 6            & 2,29,(2,4,5,4)/10 & \hbox{} \\ \hline     
I27& 13& 1 & 9 &    &  15 &(1,2,2,2)/15  &                      & I26, I27      & 2,3,4,5,6    & 5,60,(9,6,10,6)/15& \hbox{} \\ \hline       
I26& 14& 1 & 9 &    &  12 &(1,1,1,8)/15  &                      & I26, I27      & 3,8,12,14,18 & 6,43,(10,3,3,3)/15& \hbox{} \\ \hline       
I32& 16& 0 &10 &    &  15 &(0,0,0,1/5)   & I32\fold I45\fold I51&               & 3,8,12,14,18 & =                 & \hbox{ (H3) \cite{Boalch2007Highergenus}}\\ \hline   
I31& 17& 0 &10 &    &  11 &(0,0,0,3/5)   & I31\fold I44\fold I50&               & 3,8,12,14,18 & =                 & \hbox{ (H3)' \cite{Boalch2007Highergenus}}\\ \hline
I29& 18& 0 &10 &    &  18 &(3,7,7,7)/30  &                      & I29, I30      & 2,3,4,5,6    & 2,43,(3,5,5,5)/15 & \hbox{non-monic} \\ \hline    
I30& 19& 0 &10 &    &  15 &(1,1,9,1)/30  &                      & I29, I30      & 6,8,10,19,23 & 6,43,(9,5,5,5)/15 & \hbox{} \\ \hline       
I36& 22& 1 &12 &    &  23 &(1,3,3,11)/30 &                      &               & 3            & 6,82,(9,5,5,3)/15 & \hbox{} \\ \hline        
I34& 23& 1 &12 &  1 &  23 &(5,13,5,23)/60&                      & I34, I35      & 18           & 6,63,(15,6,6,10)/30 & \hbox{} \\ \hline        
I35& 24& 1 &12 &    &  25 &(1,5,11,5)/60 &                      & I34, I35      & 6            & 6,83,(15,12,12,10)/30 & \hbox{} \\ \hline        
I33& 25& 0 &12 &    &  24 &(1,7,11,17)/60&                      &               &\hbox{none}   & 2,52,(6,12,10,15)/30 & \hbox{generic} \\ \hline  
I38& 26& 1 &15 &  2 &  46 &(2,2,2,3)/15  &                      & I37, I38      & 3,8,12,14,18 & 6,104,(6,5,5,5)/15   & \hbox{Valentiner} \\ \hline       
I37& 27& 1 &15 &  1 &  34 &(1,1,1,6)/15  &                      & I37, I38      & 3,8,12,14,18 & 12,170,(12,5,5,5)/15 & \hbox{Valentiner} \\ \hline  
I40& 28& 1 &15 &    &  26 &(0,0,1,4)/15  & I40\fold I48         & I39, I40      & 8            & 9,176,(5,9,9,10)/15  & \\ \hline 
I39& 29& 1 &15 &    &  18 &(0,0,7,2)/15  & I39\fold I47         & I39, I40      & 8            & 2,37,(2,0,0,7)/15    & \cite[p 28]{Boalch2007Highergenus} \\ \hline
I41& 31& 1 &18 &    &  39 &(0,0,0,1)/3   & I41\fold I49\fold I52&               & 3,8,12,14,18 & 6,142,(2,1,1,1)/3    & \hbox{ (H3)}'' \\ \hline 
237& 32& 1 &18 &    &  54 &(1,5,5,5)/42  &                      & 237, 238, 239 & 2,3,4,5,6    & 12,264,(14,12,12,12)/21 &  \\ \hline        
238& 33& 1 &18 &  2 &  58 &(3,3,7,3)/21  &                      & 237, 238, 239 & 6,8,10,19,23 & 2,96,(3,7,3,3)/21    & \hbox{Kitaev} \\ \hline         
239& 34& 1 &18 &    &  39 &(1,1,1,17)/42 &                      & 237, 238, 239 & 3,8,12,14,18 & 12,184,(14,6,6,6)/21 &  \\ \hline 
I43& 37& 1 &20 &  1 &  67 &(3,3,7,13)/60 &                      & I42, I43      & 8            & 12,251,(18,10,10,15)/30 & \hbox{ Kitaev} \\ \hline   
I42& 38& 1 &20 &    &  62 &(1,9,9,19)/60 &                      & I42, I43      & 3            & 4,136,(6,10,10,15)/30& \\ \hline 
I46& 39& 1 &24 &  2 &  99 &(1,1,1,3)/12  &                      &               & 3,8,12,14,18 & 12,299,(3,2,2,2)/6   & \hbox{Valentiner} \\ \hline    
\hline   
007&  4& 0 & 6 &    &   8 &(1,1,5,5)/24  & O08\fold O07          &              & 7                 & 2,22,(4,6,3,6)/12 & \hbox{} \\ \hline 
T06&  3& 0 & 6 &    &   6 &(0,0,1,1)/6   & III\fold T06          &              & 2,7,8             & 1,13,(3,3,2,2)/6  & \hbox{} \\ \hline 
O10&  9& 0 & 8 &    &   9 &(0,0,1,1)/4   & IV \fold O10\fold O13 &              & 2,7,8             & 2,27,(1,2,2,1)/4  & \hbox{} \\ \hline   
O13& 30& 0 &16 &  2 &  51 &(1,1,1,1)/8   & IV \fold O10\fold O13 &              & \hbox{all}        & 8,153,(1,2,2,2)/4 & \hbox{} \\ \hline  
I28& 15& 0 &10 &  1 &  23 &(1,1,2,2)/10  & I21\fold I28          &              & 2,7,8             & 4,57,(4,5,2,5)/10 & \hbox{} \\ \hline  
O11& 21& 0 &12 &    &  21 &(0,0,1,1)/6   & III\fold O11\fold O12 &              & 2,7,8             & 4,73,((2,2,3,3)/6 & \hbox{} \\ \hline
O12& 20& 1 &12 &  1 &  28 &(1,1,1,1)/12  & III\fold O11\fold O12 &              & 2,7,8,15,16,21,22 & 6,93,(2,3,3,3)/6  & \hbox{} \\ \hline 
I44& 36& 1 &20 &    &  33 &(0,0,3,3)/10  & I31\fold I44\fold I50 & I44, I45     & 2,7,8             & 6,90,(3,0,0,3)/10 &  \\ \hline   
I50& 43&N3 &40 &  6 & 335 &(3,3,3,3)/20  & I31\fold I44\fold I50 & I50, I51     & \hbox{all}        & =                 & \cite[p 28]{Boalch2007Highergenus}\\ \hline
I45& 35& 1 &20 &    &  59 &(0,0,1,1)/10  & I32\fold I45\fold I51 & I44, I45     & 2,7,8             & 2,69,(1,0,0,1)/10 & \\ \hline       
I51& 44&N3 &40 &  2 & 275 &(1,1,1,1)/20  & I32\fold I45\fold I51 & I50, I51     & \hbox{all}        & =                 & \cite[p 27]{Boalch2007Highergenus}\\ \hline
I47& 40&H2 &30 &  2 & 134 &(2,2,7,7)/30  & I39\fold I47          & I47, I48     &2,7,8              & 7,195,(7,2,2,7)/30&  \\ \hline  
I48& 41&H2 &30 &  1 & 145 &(1,1,4,4)/30  & I40\fold I48          & I47, I48     &2,7,8              & 1,170,(1,4,4,1)/30&  \\ \hline    
I49& 42&H3 &36 &    & 153 &(0,0,1,1)/6   & I41\fold I49\fold I52 &              &2,7,8              & 6,246,(1,0,0,1)/6 & \cite[p 29]{Boalch2007Highergenus}\\ \hline
I52& 45&N7 &72 &  6 & 975 &(1,1,1,1)/12  & I41\fold I49\fold I52 &              & \hbox{all}        & =                 & \hbox{} \\ \hline
\end{array}
\nonumber
\end{equation}
}
\end{center}
\label{TableNotation}
\end{sidewaystable}
\vfill \eject

\subsection{Solutions of genus zero}
\label{sectiong0}

The chosen representatives for the genus zero are those of Table \ref{TableNotation},
but other choices allow a better symmetry of the expressions,
see Appendix \ref{sectionSimpleg0} for details.


\subsubsection{Unfolded solutions of genus zero}

\begin{itemize}
	\item II, III, IV
\begin{eqnarray}
& &             II, b=2, \vect=(a,a,b,b),    u=s, x=s^2,                         u^2-x=0,
\nonumber\\ & & III,b=3, \vect=(a,2a,a,1/3), u=s, x=\frac{s^3}{3 s-2},           u^3-3 x u +2 x=0,  
\nonumber\\ & & IV ,b=4, \vect=(b,b,b,1/2),  u=s, x=-\frac{s^2(s^2-2 s)}{2 s-1}, u^4-2 u^3 +2 x u-x=0.
\end{eqnarray}	
	
	\item I20=LT01.
\begin{eqnarray}
& &
b=5,\vect=\frac{(1,3,3,4)}{15},
x=\frac{(2s-1)^2 (2s+9)^3}{2(20s^2+27)^2}\ccomma
u=\frac{(2s+9)(2s-1)(2s-3)}{2(20s^2+27)}\cdot
\end{eqnarray}	
	
	\item I21=LT02.
\begin{eqnarray}
& &
b=5,\vect=\frac{(0,0,1,2)}{5},
x=\frac{(3s-1)^2 (s+3)^3}{8 s^3 (s+5)^2}\ccomma
u=\frac{(s+3)(3s-1)}{2 s(s+5)}\cdot
\end{eqnarray}	
		
	\item O08=LT05.
\begin{eqnarray}
& & 
b=6,\vect=\frac{(0,0,1,5)}{12},
x=\frac{(3s-8)^4}{s^4(s-3)^2}\ccomma
u=\frac{3s-8}{s(s-3)}\cdot 
\end{eqnarray}	

	\item Siblings I23=LT06, I22=LT07.
\begin{eqnarray}
& & 
b=6,
\vect_{I23}=\frac{(1,5,7,7)}{30},
\vect_{I22}=\frac{(1,1,5,13)}{30},
x=-\frac{(2s-1)(5s-1)^2}{s^3(9s-2)(9s-5)^2},
\nonumber\\ & &
u_{I23}=\frac{(5s-1)(2s-1)}{s^2(9s-5)}, 
u_{I22}=-\frac{5s-1}{s(9s-5)}\cdot 
\end{eqnarray}	

	\item Klein=LT08.
\begin{eqnarray}
& &
b=7,\vect=\frac{(1,1,2,1)}{7},
x=\frac{2(s-3)^3(s^2-6s+16)^2}{s^3(s^2-7s+14)^2}\ccomma
u=\frac{(s^2-6s+16)(s-3)^2(s-4)}{s^2(s^2-7s+14)}\cdot 
\label{eqK-rep12terms}
\end{eqnarray}

	\item O09=LT10.
\begin{eqnarray}
& &
b=8,\vect=\frac{(1, 3, 3, 7)}{24},
x=-\frac{4(4+4s+3s^2)^2}{s^3(s^2+2s+4)^2(s+4)}\ccomma
u=\frac{2(s+1)(4+4s+3s^2)}{s(s+4)(s^2+2s+4)}\cdot 
\end{eqnarray}	

	\item Siblings I24=LT11, I25=LT12. 
\begin{eqnarray}
& &
b=8,
\vect_{I24}=\frac{(1,5,1,7)}{20}, 
\vect_{I25}=\frac{(1,3,5,3)}{20},
x=\frac{(s-3)^3(s+5)^5}{64s^3(s^2-6s+25)^2}\ccomma
\nonumber\\ & &
u_{I24}=\frac{(s+5)(s-3)(s^2-10s+5)}{8s(s^2-6s+25)}\ccomma 
u_{I25}=\frac{(s-5)(s+5)^2(s-3)^2}{16s^2(s^2-6s+25)}\cdot
\end{eqnarray}	

	\item I32=LT16.
\begin{eqnarray}
& &
b=10,\vect=\frac{(0,0,0,1)}{5},
x=\frac{(s-1)^5(3s+1)^3(s^2+4s-1)}{256s^5(5s^2-1)}\ccomma
u=-\frac{(3s+1)(s-1)^3}{16s^3}\cdot
\end{eqnarray}	

	\item I31=LT17.
\begin{eqnarray}
& &
b=10,\vect=\frac{(0,0,0,3)}{5},
x=-\frac{(s^2+s-1)(2s+1)^3}{s^5(s^2-1-s)(s+2)^3}\ccomma
u=\frac{2s+1}{s(s+2)}\cdot
\end{eqnarray}	

	\item Siblings I29=LT18, I30=LT19.
\begin{eqnarray}
& &
b=10,
\vect_{I29}=\frac{(3,7,7,7)}{30},
\vect_{I30}=\frac{(1,1,9,1)}{30},
x=\frac{(2s^2-s+2)^2(s+2)^5(3s-2)}{8s^5(5s^2+12)^2}\ccomma
\nonumber\\ & &
u_{I29}=\frac{(2s^2-s+2)(s+2)^2(s^2-s+2)}{2s^3(5s^2+12)}\ccomma
u_{I30}=\frac{(2s^2-s+2)(s+2)^4}{4s^4(5s^2+12)}\cdot
\end{eqnarray}	

	\item I33=LT25.
\begin{eqnarray}
& & {\hskip -15.0 truemm}
b=12,\vect=\frac{(1,7,11,17)}{60},
x=\frac{4(7s-10)(s^2+20)^2(s^2-5s+10)^3(s-5)}{27s^5(s^2-4s+20)^2(s-4)^3}\ccomma
\nonumber\\ & &
u=\frac{2(7s-10)(s^2+20)(s^2-5s+10)}{3s^2(s^2-4s+20)(s-4)^2}\cdot
\end{eqnarray}	

\end{itemize}

\vfill\eject

\subsubsection{Folded solutions of genus zero}

The correspondence between the $s$ of the folded solution and the $s$ of the unfolded one is not rational,
but there exists a representation on $\mathbb{Q}$
for all the solutions but one
(O13=LT30, which requires the extension $i$). 

\begin{itemize}

	\item O07=LT04.
\begin{eqnarray}
& &
b=6,\vect=\frac{(1,1,5,5)}{24},
x=\frac{(s+3)^3(3s+1)^3}{4s(9s^2+14s+9)^2}\ccomma
u=\frac{(s+3)(3s+1)^2}{2(9s^2+14s+9)}\cdot
\end{eqnarray}	

	\item T06=LT03.	
\begin{eqnarray}
& &
b=6,\vect=\frac{(0,0,1,1)}{6},
x=-\frac{(s-1)^3(s-3)^3}{s^3(s-2)^3}\ccomma
u=\frac{(s-3)(s-1)^2}{s(s-2)^2}\cdot
\end{eqnarray}	

	\item O10=LT09.
\begin{eqnarray}
& &
b=8,\vect=\frac{(0,0,1,1)}{4},
x=-\frac{16(s-1)^3(s-3)^3}{s^3(s-2)^2(s-4)^3}\ccomma
u=\frac{4(s-1))(s-3)^2}{s^2(s-2)(s-4)}\cdot
\end{eqnarray}	

	\item O13=LT30. 
\begin{eqnarray}
& &
b=16,\vect=\frac{(1,1,1,1)}{8},
x=\frac{(s^2-1)^2(s^4+6s^2+1)^3}{32s^2(s^4+1)^3}\ccomma
\nonumber\\ & & 
u=\frac{(1+i)(s^2+(1-i)s+i)(s^2+2is+1)(s^2-1)(s^2-2is+1)^2}{8s(s^2-i)^2(s^2+i)(s^2+(1+i)s-i)}\cdot
\end{eqnarray}	
This representative is invariant under the $4!$ homographies.
This is the only solution without any real branch \cite[p 99]{Boalch2007Bolibrukh}.
Its representation depends on one algebraic number, normalized to $i$.

	\item I28=LT15.
\begin{eqnarray}
& & {\hskip -15.0 truemm}
b=10,\vect=\frac{(1,1,2,2)}{10},
x=-\frac{(2s+3)^3(5s^2+4s+1)^2} {s^3(s^2+4s+5)^2(3s+2)^3}\ccomma
u= \frac{(2s+3)(s-1)(5s^2+4s+1)}{s  (s^2+4s+5)  (3s+2)^2(s+1)}\cdot
\end{eqnarray}	

	\item O11=LT21.
\begin{eqnarray}
& &
b=12,\vect=\frac{(0,0,1,1)}{6},
x=\frac{(6s^2-8s+3)^2(2s-3)^4}{16s^4(2s^2-8s+9)^2(s-1)^4}\ccomma
u=\frac{(2s-3)(6s^2-8s+3)}{4s(2s^2-8s+9)(s-1)^3}\cdot
\end{eqnarray}	
	
\end{itemize}

\vfill\eject
	
\subsection{Solutions of genus one}
\label{sectiong1}

The elliptic curve
$t^2=c_3 s^3+c_2 s^2+c_1 s + c_0, c_3 \not=0$,
an affine transform of the curve of Weierstrass
${\wp'(\lambda)}^2=4 \wp(\lambda)^3 - g_2 \wp(\lambda) -g_3=4(\wp(\lambda)-e_1)(\wp(\lambda)-e_2)(\wp(\lambda)-e_3)$,
is chosen so as to minimize the size of the integer numbers.

Each solution of genus $g$ higher than or equal to one admits a representative 
invariant under the involution $(u,x) \to (1-u,1-x)$, 
this representative therefore admits for $g=1$ the representation \cite{Boalch-Icosa},
\begin{eqnarray}
& &
x=\frac{1}{2}+R_1(s) t,
u=\frac{1}{2}+R_2(s) t.
\end{eqnarray}

This representative 
may not be the one in Table \ref{TableNotation}
(chosen to minimize the degree $d$ and the number of terms of the algebraic curve),
but they differ by a homography.

This convention of simplicity is detailed in the Appendix \ref{sectionSimpleg1}.

\vfill\eject
\subsubsection{Unfolded solutions of genus one}

All the representatives in this section are chosen invariant under the
involution $(x,u,s,t) \to (1-x,1-u,s,-t)$.

\begin{itemize}

	\item Siblings I27=LT13, I26=LT14.
\begin{eqnarray}
& &
b=9,\vect_{I27}=\frac{(1,2,2,2)}{15}\ccomma \vect_{I26}=\frac{(1,1,1,8)}{15}\ccomma
t^2=\frac{s(2s+1)(5s+16)}{36}\ccomma 
s=\frac{12\wp(\lambda)-37}{30}\ccomma t=\frac{\wp'(\lambda)}{15}\ccomma
\nonumber\\ & & 
x=\frac{1}{2}-\frac{25s^4-40s^3-84s^2-136s-8}{72s(2s+1)^2} t,
u_{I27}=\frac{1}{2}+\frac{5s^2+2s+2}{6s(2s+1)} t, 
u_{I26}=\frac{1}{2}-\frac{s-1}{s(2s+1)} t. 
\end{eqnarray}	
All the poles $s$ being also zeroes of $t$,
the fields $x$, $u_{I27}$, $u_{I26}$ are polynomials of the elliptic fonctions of Jacobi.

	\item I36=LT22. 
The elliptic curve is identical to that of the siblings (I42, I43). 
\begin{eqnarray}
& & 
b=12,\vect=\frac{(1,3,3,11)}{30}\ccomma
t^2=3(5s-2)(16s^2-25s+10), 
\nonumber\\ & & 
s=\frac{\wp(\lambda)+157}{240}\ccomma  t=\frac{\wp'(\lambda)}{480}\ccomma g_2=-7788, g_3=432856,
\nonumber\\ & & {\hskip -15.0 truemm} 
x=\frac{1}{2}-\frac{160s^6+600s^5-2865s^4+4100s^3-2820s^2+960s-128}{6s^2(16s^2-25s+10)t}\ccomma
u=\frac{1}{2}-\frac{31s^2-46s+16}{2st}\cdot
\end{eqnarray}	

	\item Siblings I34=LT23, I35=LT24.	
\begin{eqnarray}
& &
b=12,
\vect_{I34}=\frac{(13,5,5,23)}{60}\ccomma  
\vect_{I35}=\frac{(1,5,5,11)}{60}\ccomma  
\nonumber\\ & & 
t^2=s(32s^2-95s+80), s=\frac{3\wp(\lambda)+95}{96}\ccomma t=\frac{\wp'(\lambda)}{64}\ccomma
\nonumber\\ & & 
x=\frac{1}{2}+\frac{2048s^6-15360s^5+48000s^4-77840s^3+65685s^2-20328s-4000}{54(32s^2-95s+80)t}\ccomma
\nonumber\\ & & 
u_{I34}=\frac{1}{2}-\frac{64s^3-216s^2+249s-200}{6(8s-13)t}\ccomma
u_{I35}=\frac{1}{2}+\frac{64s^3-288s^2+447s-200}{18t}\cdot
\end{eqnarray}	
The fields $x$ and $u_{I35}$ are polynomials of the fonctions of Jacobi. 

\item Siblings	I38=LT26, I37=LT27. 
\begin{eqnarray}
& &
b=15,
\vect_{I38}=\frac{(2,2,2,3)}{15}\ccomma 
\vect_{I37}=\frac{(1,1,1,6)}{15}\ccomma 
d_{I38}-b=2, d_{I37}-b=1, 
\nonumber\\ & & 
t^2=3(5s-2)(4s^2-5s+10), s=\frac{3\wp(\lambda)+11}{20}\ccomma t=\frac{9\wp'(\lambda)}{40}\ccomma
\nonumber\\ & & 
x=\frac{1}{2}-\frac{50s^7-140s^6+438s^5-490s^4+655s^3+1290s^2-640s+1024}{486s^2(4s^2-5s+10)^2} t,
\nonumber\\ & & 
u_{I38}=\frac{1}{2}-\frac{10s^4-22s^3+51s^2-22s+64}{54s(s+2)(4s^2-5s+10)}t,
u_{I37}=\frac{1}{2}+\frac{10s^3-3s^2+30s-64}{6st}\cdot
\end{eqnarray}		
 
\item	Siblings I40=LT28, I39=LT29.  
\cite[p 208]{Boalch-Icosa} 
\cite[p 28]{Boalch2007Highergenus}. 
\begin{eqnarray}
& &
b=15,
\vect_{I39}=\frac{(2,0,0,7)}{15}\ccomma  
\vect_{I40}=\frac{(1,0,0,4)}{15}\ccomma  
d_{I39}-b=2, d_{I40}-b=1,
\nonumber\\ & &  
t^2=3 (s+5)(4s^2+15s+15), s=\frac{\wp(\lambda)-35}{12}\ccomma  t=\frac{\wp'(\lambda)}{24}\ccomma  
\nonumber\\ & &  
x=\frac{1}{2}-\frac{2s^7+10s^6-90s^4-135s^3+297s^2+945s+675}{18(4s^2+15s+15)^2(s^2-5)}t,
\nonumber\\ & & 
u_{I39}=\frac{1}{2}-\frac{(2s^2+3s-3)}{6(s+1)(4s^2+15s+15)}t,
u_{I40}=\frac{1}{2}-\frac{2s^3+4s^2-9s-15}{2t}\cdot
\end{eqnarray}		

\item I41=LT31=(H3)''. 
\cite[p 29]{Boalch2007Highergenus}. 
\begin{eqnarray}
& &
b=18,\vect=\frac{(0,0,0,1)}{3}\ccomma 
t^2=s(8s^2-11s+8), s=\frac{12 \wp(\lambda)+11}{24}\ccomma t=\frac{\wp'(\lambda)}{2}\ccomma
\nonumber\\ & & 
x=\frac{1}{2}+\frac{(s+1)(32s^8-320s^7+1112s^6-2420s^5+3167s^4-2420s^3+1112s^2-320s+32)}{54s^2(s-1)(8s^2-11s+8)t}\ccomma
\nonumber\\ & &  
u=\frac{1}{2}-\frac{8s^3-12s^2+3s-4}{6t}\cdot
\end{eqnarray}	
This representative is also invariant under the involution $(x,s,t) \to (x,1/s,-t/s^2)$.

\item	Siblings 237=LT32, 238=LT33, 239=LT34. 
\begin{eqnarray}
& &
b=18,
\vect_{LT32}=\frac{(1,5,5,5)}{42}\ccomma 
\vect_{LT33}=\frac{(3,3,3,7)}{21}\ccomma 
\vect_{LT34}=\frac{(1,1,1,17)}{42}\ccomma 
\nonumber\\ & & 
t^2=(2s-1)(4s^2-2s+7), s=\frac{\wp(\lambda)+6}{18}\ccomma t=\frac{\wp'(\lambda)}{54}\ccomma
\nonumber\\ & & 
x=\frac{1}{2}-\frac{16s^9-72s^8+144s^7-336s^6+252s^5-504s^4-294s^3+225s^2-288s+128}{54s^2(4s^2-2s+7)t}\ccomma
\nonumber\\ & & 
u_{LT32}=\frac{1}{2}+\frac{8s^5-4s^4+20s^3-8s^2-5s+16}{18st}\ccomma
\nonumber\\ & & 
u_{LT33}=\frac{1}{2}+\frac{4s^4-4s^3+12s^2-s+16}{6s(2s+1)(4s^2-2s+7)}t,
\nonumber\\ & &
u_{LT34}=\frac{1}{2}-\frac{4s^3+3s-16}{6st}\cdot
\end{eqnarray}

\item 
Siblings I43=LT37, I42=LT38. 

 The curve $(s,t)$ is identical to that of I36. 
\begin{eqnarray}
& &
b=20,
\vect_{I43}=\frac{(7,3,3,13)}{60}\ccomma 
\vect_{I42}=\frac{(1,9,9,19)}{60}\ccomma 
\nonumber\\ & & 
t^2=3s(16s^2-61s+64), 
s=\frac{\wp(\lambda)+61}{48}\ccomma t=\frac{\wp'(\lambda)}{96}\ccomma g_2=-7788, g_3=432856,
\nonumber\\ & & 
x=\frac{1}{2}+\frac{P(s)}{6(16s^2-61s+64)(2s^2-6s+5)^2t}\ccomma
\nonumber\\ & & 
P(s)=512s^{10}-7680s^9+51840s^8-206560s^7+535380s^6-935448s^5
\nonumber\\ & &  \phantom{12345}   
         +1098280s^4-825660s^3+343875s^2-41120s-13824,  
\nonumber\\ & & 
u_{I43}=\frac{1}{2}-\frac{32s^5-216s^4+590s^3 -846s^2 +689s-288}{2(4s-7)(2s^2-6s+5)t}\ccomma                       
\nonumber\\ & & 
u_{I42}=\frac{1}{2}+\frac{32s^5-256s^4+826s^3-1322s^2+1023s-288}{2(2s^2-6s+5)t}\cdot
\end{eqnarray}		

\item	I46=LT39. \cite[p.~213]{Boalch-Icosa}.
\begin{eqnarray}                      
& &
b=24,d=b+2,\vect=\frac{(1,1,1,3)}{12}\ccomma 
t^2=(s+2)(8s^2-7s+2), s=\frac{\wp(\lambda)}{2}-\frac{3}{8}\ccomma t=\frac{\wp'(\lambda)}{2}\ccomma
\nonumber\\ & & 
x=\frac{1}{2}+\frac{(s^2+4s-2)P(s)}{2(s+2)^2(3s^2-2s+2)^2(8s^2-7s+2)t},
\nonumber\\ & & 
P(s)=8s^{10}+16s^9+24s^8-84s^7+429s^6-312s^5+258s^4-288s^3+288s^2-128s+32,
\nonumber\\ & &
u=\frac{1}{2}-\frac{4s^6+16s^5+9s^4-2s^3-34s^2+24s-8}{2(3s^2-2s+2)(s-2)t}\cdot
\end{eqnarray}	

\end{itemize}

\vfill\eject
\subsubsection{Folded solutions of genus one}

\begin{itemize}
	\item 
Siblings I44=LT36 et I45=LT35. 
Transport of $\vect_{I44}=(3,0,0,3)/10$ and of $\vect_{I45}=(1,0,0,1)/10$ \cite[\S 4.1]{Boalch2007Highergenus}.
\begin{eqnarray}
& &
b=20, \vect_{I44}=\frac{(0,3,3,0)}{10}\ccomma  \vect_{I45}=\frac{(0,1,1,0)}{10}\ccomma  
%
%
t^2=3(s-1)(5s^2+5s-1), s=\frac{4}{15}\wp(\lambda)\ccomma t=\frac{4}{15}\wp'(\lambda)\ccomma 
\nonumber\\ & &
x=\frac{1}{2}-i \frac{P(s)}{576(5s^2-10s-4)(5s^2+5s-1)(s-1)^2t}\ccomma
\nonumber\\ & & 
P(s)=3125s^{10}-12500s^9+48000s^7-35400s^6-117936s^5
\nonumber\\ & &  \phantom{1234567} 
     +191760s^4+27840s^3-58320s^2+10240s+2240,
\nonumber\\ & & 
u_{I44}=\frac{1}{2}+i \frac{(5s^2-22s+26)(5s^2+2s+2)^2(5s-2)(s+2)}{144(5s^2-10s-4)(5s^2+5s-1)(s-1)^2}
                   +i \frac{(25s^3-30s^2-42s+20)^2+(54s)^2}{120(5s^2-10s-4)(s-1)t}\ccomma
\nonumber\\ & & 
u_{I45}=\frac{1}{2}+i \frac{(5s^2-22s+26)(5s^2+2s+2)(5s-2)(s+2)}{24(s-1)(5s^2-10s-4)(5s^2+5s-1)} 
                   +i \frac{5s^2-4s+8}                                {2(5s^2-10s-4)(5s^2+5s-1)}t. 
\end{eqnarray}

This representative is invariant under the involution $(x,u,s,i,t) \to (1-x,1-u,s,-i,t)$.

\textit{Remark}.
This representation is obtained by folding 
the genus zero solutions I31 and I32 \cite[\S 4.1]{Boalch2007Highergenus},
which defines the elliptic curve $T^2=(9S^2-2S+9)(S^2-2S+17)$,
then by the homography which sends to infinity one of the four zeroes of $T$ 
(for instance $1+4 i$),
\begin{eqnarray}
& &
S=1+4i+\frac{18(1+2i)}{X-\frac{3}{4}(11-3i)}\ccomma
T=\frac{48i(1+2i)Y}{[X-\frac{3}{4}(11-3i)]^2}\ccomma
Y^2=\frac{4X-15}{16X^2+60X-45}\cdot
\end{eqnarray}
The reason why the Cremona transformation \cite[\S 4.1]{Boalch2007Highergenus}
\begin{eqnarray}
& &
S=\frac{t-9s-81}{t+3s-9}\ccomma  T=16\frac{t^2+54 t+18s^2+540s+405}{(t+3s-9)^2}\ccomma 
t^2=s^3-270s-675,
\end{eqnarray}
does not display the invariance $(x,u) \to (1-x,1-u)$  
is its independence on the number $i$.

\item 
O12=LT20. Transport of $\vect=(2,3,3,3)/6$ \cite[p 99]{Boalch2007Bolibrukh}.
\begin{eqnarray}
& &
b=28,d=b+1,\vect=\frac{(1,1,1,1)}{12}\ccomma 
t^2=(2s+1)(9s^2+2s+1),s=\frac{3\wp(\lambda)-13}{54}\ccomma t=\frac{\wp'(\lambda)}{36}\ccomma
\nonumber\\ & & 
x=\frac{1}{2}+\frac{27s^4+28s^3+26s^2+12s+3)s}{(s+1)^3(9s^2+2s+1)^2}t\ccomma
u=\frac{1}{2}+\frac{11s^3+5s+1+7s^2}{2(s+1)^2t}\cdot
\end{eqnarray}		

\end{itemize}


\vfill\eject
\subsection{Folded solutions of genus higher than one}

\begin{itemize}
	\item Siblings I47, I48 (genus two, hyperelliptic).
	
Transport of $\vect_{I47}=(7,2,2,7)/30$ ($\vect_{I48}=(1,4,4,1)/30$ is conserved) \cite[\S 4.3]{Boalch2007Highergenus}. 
\begin{eqnarray}
& &
b=40, d_{I47}-b=2, d_{I48}-b=1, \vect_{I47}=\frac{(2,7,7,2)}{30}\ccomma  \vect_{I48}=\frac{(1,4,4,1)}{30}\ccomma  
\nonumber\\ & & 
t^2=(3s+1)(s+3)(s^2+1)(3s^2+4s+3),
\nonumber\\ & & 
x=\frac{1}{2}+\frac{P(s)}{54}t,
\nonumber\\ & &  
P(s)=81s^{14}+270s^{13}+567s^{12}+540s^{11}+621s^{10}+1314s^9+2955s^8
\nonumber\\ & &  \phantom{12345} 
+3688s^7+2955s^6+1314s^5+621s^4+540s^3+567s^2+270s+81,
\nonumber\\ & & 
u_{I47}=\frac{1}{2}-\frac{4s^2(3s^2+3s+2)(2s^2+3s+3)(9s^4-2s^2+9)}{9(s-1)(3s^2+4s+3)(s^2+1)^2(s+1)^3(3s^2+2s+3)}
\nonumber\\ & &  \phantom{12345} 
                   +\frac{(3s^3+3s^2+s-3)(3s^3-s^2-3s-3)}{6(s+1)(3s^2+4s+3)(s^2+1)^2(3s^2+2s+3)}t,
\nonumber\\ & & 
u_{I48}=\frac{1}{2}+\frac{27s^9+63s^8+108s^7+36s^6+42s^5+130s^4+300s^3+228s^2+99s-9}{18(s-1)(s+1)^4(s^2+1)^2(3s+1)(3s^2+4s+3)}t.
\end{eqnarray}

The involution $(s,t) \to (1/s,t/s^3)$ leaves invariant $x$, $u_{I48}$ and the third term of $u_{I47}$,
it changes the sign of the second term of $u_{I47}$.

	\item I49 (genus three, hyperelliptic).
				
				
Transport of $\vect=(1,0,0,1)/6$ \cite[\S 4.4]{Boalch2007Highergenus} \cite{LT2014}. 
\begin{eqnarray}
& &
b=36, \vect=\frac{(0,1,1,0)}{6}\ccomma  
t^2=\frac{1}{75}(s^2+2s+5)(s^2+4s+5)(3s^4+30s^3+110s^2+150s+75),  
\nonumber\\ & & 
x=\frac{1}{2}-\frac{5 (s+1) (5+s) P(s)}{6(s^2-5) (3 s^4+30 s^3+110 s^2+150 s+75)^2 (s^2+2 s+5)^3 (s^2+4 s+5)^3}t,
\nonumber\\ & & 
P(s)=27 s^{16}+648 s^{15}+7452 s^{14}+53568 s^{13}+266292 s^{12}+968400 s^{11}+2714980 s^{10}+6371400 s^9
\nonumber\\ & &  \phantom{12345} 
+14138050 s^8+31857000 s^7+67874500 s^6+121050000 s^5 +166432500 s^4+167400000 s^3
\nonumber\\ & &  \phantom{12345} 
+116437500 s^2+50625000 s+10546875,
\nonumber\\ & & 
u=\frac{1}{2}-\frac{2s(s^2+5s+10)(3s^2+10s+15)(2s^2+5s+5)(3s+5)^2(3+s)^2}
                         {3(s^2-5)(3s^4+30s^3+110s^2+150s+75)(s^2+2s+5)^2(s^2+4s+5)}
\nonumber\\ & &  \phantom{12345} 
								   -\frac{5(s^3+25s^2++75s+75)(3s^3+15s^2+25s+5)}
									       {2(s^2-5)(3s^4+30s^3+110s^2+150s+75)(s^2+2s+5)^2}t.
\end{eqnarray}
This representative is invariant under the involution $(x,u,s \sqrt{5},t) \to (1-x,1-u,1/(s \sqrt{5}),(t/25)(5/s)^4)$.

	\item Siblings I50, I51 (genus three, non-hyperelliptic).
	
The previously chosen representatives \cite[\S 4.2]{Boalch2007Highergenus} are minimal
and invariant under the $4!$ homographies.

   The folding of the siblings (I44, I45) generates a representation by two elliptic curves,
then the homography $s \to 1+4/s$ simplifies the expressions of \cite[\S 4.2]{Boalch2007Highergenus},
\begin{eqnarray}
& &
b=40, d_{I50}-b=6, d_{I51}-b=2, \vect_{I50}=\frac{(3,3,3,3)}{20}\ccomma  \vect_{I51}=\frac{(1,1,1,1)}{20}\ccomma  
\nonumber\\ & & 
t_1^2=-s(2s-1)(s+2),
t_2^2=(2s-1)(s^2+2s+5),
\nonumber\\ & & 
x=\frac{1}{2}-\frac{s^{10}+10s^9+45s^8+120s^7+190s^6-4s^5-410s^4-680s^3+25s^2+90s-27}{16s^2(s^2+2s+5)(s+2)^3(2s-1)^2}t_1,
\nonumber\\ & &  
u_{I50}=\frac{1}{2}-\frac{(s^2+4s-1)(s^2+4s+9)(s^2+1)^2}{4s(s^3+3s^2+15s+1)(s+2)t_2}-\frac{s^3+3s^2+3s-3}{2(s^3+3s^2+15s+1)s}t_1\ccomma
\nonumber\\ & & 
u_{I51}=\frac{1}{2}-\frac{(s^2+4s-1)(s^2+4s+9)(s^2+1)}{4s(s+1)(s+2)^2t_2}-\frac{s-3}{2s(s+1)(s+2)^2}t_1\cdot
\end{eqnarray} 
The two elliptic curves admit a rational representation
in terms of the genus three curve, whose minimal degree is four \cite[\S 4.2]{Boalch2007Highergenus},
\begin{eqnarray}
& &
5(p^4+q^4)+6(p^2q^2+p^2+q^2)+1=0,
\nonumber\\ & & 
s=-\frac{2p^2}{p^2+q^2+1}\ccomma t_1=p t_2, t_2=-\frac{4(q^2+1))q}{(p^2+q^2+1)^2}\cdot.
\end{eqnarray}
 
	\item 
	I52 (genus seven, non-hyperelliptic).
				
   The folding of I49 generates a representation by two hyperelliptic curves,
and the homography $s \to -1+3/s$	
simplifies the representation of Ref.~\cite[\S 4.5]{Boalch2007Highergenus},
\begin{eqnarray}
& &
b=72, d_{I52}-b=6, \vect_{I52}=\frac{(1,1,1,1)}{12}\ccomma  
\nonumber\\ & & 
t_1^2=s(3s-2)(2s-3)(2s^2-s+2)(4s^2-7s+4),
t_2^2=-s^3(3s^2-4s+3)        (4s^2-7s+4),
\nonumber\\ & & 
x=\frac{1}{2}-\frac{(s^2-1)P(s)}{s(3s-2)(2s-3)(3s^2-4s+3)t_1^3},
\nonumber\\ & & 
P(s)=864(s^{16}+1)-10368(s^{15}+s) +59616(s^{14}+s^2)-221184(s^{13}+s^3)+599976(s^{12}+s^4)
\nonumber\\ & &  \phantom{1234} 
     -1263960(s^{11}+s^5)+2127908(s^{10}+s^6)-2899008(s^9+s^7)+3212357s^8,
\nonumber\\ & & 
u_{I52}=\frac{1}{2}+\frac{3(2s^2-2s+1)^2(s^2-3s+1)(s^2-2s+2)(6s^4-6s^3+s^2-6s+6)s}{(3s-2)^2(2s-3)(2s^2-s+2)(2s^3-4s^2+6s-3)t_2}
\nonumber\\ & &  \phantom{1234567} 
         -\frac{s(6s^3-12s^2+8s+1)}{2(3s-2)^2        (2s^2-s+2)(2s^3-4s^2+6s-3)}t_1,
\end{eqnarray}
because it then gains the involution $(x,s,t_1) \to (x,1/s,-t_1/s^4)$.
This representative is invariant under the $4!$ homographies.

The two hyperelliptic curves admit a rational representation 
in terms of the genus seven curve, whose minimal degree is eight \cite[\S 4.5]{Boalch2007Highergenus},
\begin{eqnarray}
& &   
9(q^6p^2+q^2p^6)+
18q^4p^4+
4(q^6+p^6)+
26(q^4p^2+q^2p^4)+
8(q^4+p^4)
\nonumber\\ & &  \phantom{1234} 
+
57q^2p^2+
20(q^2+p^2)+
16=0,
\nonumber\\ & & 
p^2+q^2=S, p^2-q^2=D, (3S^2+5S+8)^2-(9S^2+14S+41)D^2=0,
\nonumber\\ & & 
s=\frac{3(S-1)}{4(S+1)}+\frac{(9S^2+14S+41)D}{4(S+1)(3S^2+5S+8)}\ccomma
\nonumber\\ & & 
t_1=\frac{(9S^2+14S+41)pq}{8(S+1)^4}\left[3(S-17)+\frac{41S^2+46S+329)D}{3S^2+5S+8}\right]\ccomma
\nonumber\\ & & 
t_2=\frac{15S^3-43S^2-23S-397}{8q(S+1)^3}+\frac{(5S^2-4S+63)(9S^2+14S+41)D}{8q(3S^2+5S+8)(S+1)^3}\cdot
\end{eqnarray}

\end{itemize}

\appendix
\section{Transformations conserving $\PVI$}

The Table \ref{TablehomographiesPVI} \cite[p.~315]{CMBook2}
lists the 24 homographies.

\begin{table}[ht] 
\caption[The 24 homographies which conserve $\PVI$]{
The 24 homographies $(u,x) \to (U,X)$ which conserve $\PVI$,
ordered by increasing values of the order of the homography.
The numbering in the first column is that of Ref.~\cite{GLBook}.
}
\vspace{0.2truecm}
\begin{center}
\begin{tabular}{| c | c | c | l | l |}
\hline 
num&order& $\infty 0 1 x$ & independent var & dependent var
\\ \hline \hline 
 1 &  1  & $\infty 0 1 x$ &  $x=X$       & $u=U$                \\ \hline \hline 
 7 &  2  & $0 \infty x 1$ &  $x=X$       & $u=X/U$              \\ \hline 
15 &  2  & $1 x \infty 0$ &  $x=X$       & $u=(U-X)/(U-1)$      \\ \hline 
22 &  2  & $x 1 0 \infty$ &  $x=X$       & $u=X(U-1)/(U-X)$     \\ \hline \hline 
 6 &  2  & $\infty x 1 0$ &  $x=X/(X-1)$ & $u=(U-X)/(1-X)$      \\ \hline 
11 &  2  & $0 1 x \infty$ &  $x=X/(X-1)$ & $u=X(1-U)/((1-X) U)$ \\ \hline 
18 &  2  & $1 0 \infty x$ &  $x=X/(X-1)$ & $u=U/(U-1)$          \\ \hline 
20 &  2  & $x \infty 0 1$ &  $x=X/(X-1)$ & $u=-X/(U-X)$         \\ \hline \hline 
 2 &  2  & $\infty 0 x 1$ &  $x=1/X$     & $u=U/X$              \\ \hline 
 8 &  2  & $0 \infty 1 x$ &  $x=1/X$     & $u=1/U$              \\ \hline \hline 
 3 &  2  & $\infty 1 0 x$ &  $x=1-X$     & $u=1-U$              \\ \hline  
23 &  2  & $x 0 1 \infty$ &  $x=1-X$     & $u=(1-X)U/(U-X)$     \\ \hline \hline 
 4 &  3  & $\infty 1 x 0$ &  $x=1/(1-X)$ & $u=(1-U)/(1-X)$      \\ \hline 
10 &  3  & $0 x 1 \infty$ &  $x=1/(1-X)$ & $u=(U-X)/((1-X) U)$  \\ \hline 
14 &  3  & $1 \infty 0 x$ &  $x=1/(1-X)$ & $u=1/(1-U)$          \\ \hline 
24 &  3  & $x 0 \infty 1$ &  $x=1/(1-X)$ & $u=U/(U-X)$          \\ \hline \hline 
 5 &  3  & $\infty x 0 1$ &  $x=1-1/X$   & $u=1-U/X$            \\ \hline 
12 &  3  & $0 1 \infty x$ &  $x=1-1/X$   & $u=1-1/U$            \\ \hline
17 &  3  & $1 0 x \infty$ &  $x=1-1/X$   & $u=(X-1)U/(X(U-1))$  \\ \hline 
19 &  3  & $x \infty 1 0$ &  $x=1-1/X$   & $u=(1-X)/(U-X)$      \\ \hline \hline
16 &  4  & $1 x 0 \infty$ &  $x=1/X$     & $u=(U-X)/(X(U-1))$   \\ \hline 
21 &  4  & $x 1 \infty 0$ &  $x=1/X$     & $u=(U-1)/(U-X)$      \\ \hline \hline
 9 &  4  & $0 x \infty 1$ &  $x=1-X$     & $u=1-X/U$            \\ \hline
13 &  4  & $1 \infty x 0$ &  $x=1-X$     & $u=(1-X)/(1-U)$
\\ \hline 
\end{tabular}
\end{center}
\label{TablehomographiesPVI}
\end{table}

The unique birational transformation between 
$\PVI(u,x,\vect)$ and $\PVI(U,X,\vecT)$ is the involution defined by \cite{Okamoto1986I,CM2001c},
\begin{eqnarray}
& & 
\pmatrix{\theta_\infty \cr
         \theta_0      \cr
         \theta_1      \cr
         \theta_x      \cr}
= \frac{1}{2}               \pmatrix{ 1 & -1 & -1 & -1 \cr
                                     -1 &  1 & -1 & -1 \cr
                                     -1 & -1 &  1 & -1 \cr
                                     -1 & -1 & -1 &  1 \cr}
\pmatrix{\Theta_\infty \cr
         \Theta_0      \cr
         \Theta_1      \cr
         \Theta_x      \cr}
              + \frac{1}{2} \pmatrix{ 1 \cr 1 \cr 1 \cr 1 \cr}\ccomma
\label{eqT6Affine}
\\ & &
\frac{N}{u-U}
 =\frac{x (x-1) U'}{U (U-1)(U-x)}+\frac{\Theta_0}{U}+\frac{\Theta_1}{U-1}+\frac{\Theta_x-1}{U-x}
\\ & & \phantom{\frac{N}{u-U}}
 =\frac{x (x-1) u'}{u (u-1)(u-x)}+\frac{\theta_0}{u}+\frac{\theta_1}{u-1}+\frac{\theta_x-1}{u-x}\ccomma
\label{eqTP6uvecT}
\\ & &
N = 1-\Theta_\infty-\Theta_0-\Theta_1-\Theta_x = (1/2) \sum (\theta_j - \Theta_j).
\end{eqnarray}


The unique folding transformation between
$\PVI(u,x,\vect)$ et $\PVI(U,X,\vecT)$,
found by Kitaev \cite{Kitaev1991}, 
has been interpreted by Manin \cite{Manin1998} as a Landen transformation  
for the elliptic representation of $\PVI$.
It can be written as \cite[\S 3.2]{TOS}, 
\begin{eqnarray}
& & {\hskip -15.0 truemm}
\left\lbrace 
\begin{array}{ll}
\displaystyle{
x=\left(\frac{X^{-1/4}+X^{1/4}}{2}\right)^2,\
u=\left(\frac{X^{-1/4} U^{1/2}+X^{1/4}U^{-1/2}}{2}\right)^2,\
}\\ \displaystyle{
\forall (\lambda_1,\lambda_2):\ 
\vecT=(\lambda_1,\lambda_1,\lambda_2,\lambda_2),\
\vect=(2 \lambda_1,0,0, 2 \lambda_2).
}
\end{array}
\right.
\label{eqP6Folding2}
\end{eqnarray}

The quartic transformation \cite[Eqs.~(3.11)--(3.13)]{TOS},
\begin{eqnarray}
& & {\hskip -15.0 truemm}
\left\lbrace 
\begin{array}{ll}
\displaystyle{
x=X,\
u=\frac{(U^2-X)^2}{4 U(U-1)(U-X)}\ccomma
}\\ \displaystyle{
\forall \lambda:\ 
\vecT=(\lambda,\lambda,\lambda,\lambda),\
\vect=(4 \lambda,0,0,0),
}
\end{array}
\right.
\label{eqP6Folding4}
\end{eqnarray} 
is essentially the square \cite[Eq.~(2.2)]{TOS} of the transformation (\ref{eqP6Folding2}).

\section{Optimal representations }
\label{sectionSimple}

Depending on its genus $g$, 
each solution (except the three non-hyperelliptic I50, I51, I52) can be represented
by two rational functions $R_*$,
\begin{eqnarray}
& &
(g=0)\ x=R_1(s), u=R_2(s),
\end{eqnarray}
or by four rational functions $R_*$ and one polynomial $P$,
\begin{eqnarray}
& &
(\hbox{elliptic or hyperelliptic})\ x=R_1(s) + R_2(s) t, u=R_3(s) + R_4(s) t, t^2=P_{2 g+1}(s),
\end{eqnarray}
and an important practical question is to minimize the volume of these expressions.

Such a minimization has already been done 
mainly by Boalch with some improvements by Lisovyy and Tykhyy,
but it dealt with representatives whose gap $d-b$ is sometimes high
(see the case $d-b=12, b=18$ in the unique set ``237, 238, 239'' of three siblings elements,
Table \ref{TableNotation}).
We therefore put our effort on the minimal representatives.

A first lowering of this volume consists in chosing 
the arbitrary parameter $s$ 
so as to move the pole of $x(s)$ of maximal order to the origin.

Additional criteria allow one to obtain an even more compact representation \cite{Boalch-Icosa}.
If the equivalence class contains a representative whose curve $P(u,x)=0$
is invariant under the involution $(x,u) \to (1-x,1-u)$ resp.~$(x,u) \to (1/x,1/u)$ 
(respective numbers 3 and 8 in Table \ref{TablehomographiesPVI},
see column ``homographies'' of Table \ref{TableNotation}),
then there exists a choice of the parameter $s$ 
making $(x-1/2,u-1/2)$ odd in $s$ (resp.~$(x,u)$. 
Let us make these criteria more precise for $g=0$ and $g=1$.

\subsection{Rational representations (genus zero)}
\label{sectionSimpleg0}

The Klein solution, already representable by (\ref{eqK-rep12terms}) 
(criterium of a minimal number of terms of $P(u,x)$),
is equally representable by
\begin{eqnarray}
& &
b=7,\vect=\frac{(2,1,1,1)}{7}\ccomma 
u=\frac{1}{2}+\frac{3s^4+4s^2+9}         { s  (s^2+7)(s^2+3)}\ccomma 
x=\frac{1}{2}+\frac{7s^6+14s^4+63s^2+108}{2s^3(s^2+7)^2}\ccomma
\label{eqK-repHomog3}
\end{eqnarray}
(criterium of invariance under the involution $(x,u,s) \to (1-x,1-u,-s)$) 
or by (cf.~\cite[p 171 Eq (7)]{Boalch2005Klein}),
\begin{eqnarray}
& &
b=7,\vect=\frac{(1,1,2,1)}{7}, 
u=\frac{s(s^2+s+2)(2s+1)^2}{(2s^2+s+1)(s+2)^2}\ccomma x=\frac{(2s+1)^3(s^2+s+2)^2}{(2s^2+s+1)^2(s+2)^3}\ccomma 
\label{eqK-repHomog8}
\end{eqnarray}
(criterium of invariance under the involution  $(x,u,s) \to (1/x,1/u,1/s)$). 

The degree being the same (here $d=8$), 
the representation (\ref{eqK-repHomog3}) (which creates a parity in $s$ and therefore reduces the number of terms)
is in principle twice less voluminous
than (\ref{eqK-repHomog8})
(which exchanges numerators and denominators without reducing the number of terms).

\subsection{Eelliptic representations (genus one)}
\label{sectionSimpleg1}

There exist three main representations of an elliptic fonction of $\lambda$. 
\begin{enumerate}
	\item Sum of derivatives of fonctions $\zeta(\lambda-\lambda_j)$ of Weierstrass
	(partial fraction decomposition (``d\'ecomposition en \'el\'ements simples'') of Hermite \cite{Hermite-sum-zeta});
	\item Product of integer powers (of both signs) of fonctions $\sigma(\lambda-\lambda_j)$ of Weierstrass;
	\item Rational function of $\wp(\lambda)$ and $\wp'(\lambda)$.
\end{enumerate}

The third one is practically the simplest one but it lacks unicity
because of the addition formula of $\wp$
(the two others are insensitive to a translation of $\lambda$).

In order to minimize the size of the fractions (degrees and number of terms),
it is necessary to perform a translation of $\lambda$ which moves to the origin
the pole of $x$ of maximal order, like in the following example.

Consider an elliptic fonction  
with one pole of order one and one pole of order three,
defined par the Hermite decomposition 
\begin{eqnarray}
& &
\left\lbrace
\begin{array}{ll}
\displaystyle{
E(\lambda)=T_0''+5 (T_0-T_2),
T_0=\zeta(\lambda), T_1=\zeta(\lambda+a)-\zeta(a), T_2=\zeta(\lambda+b)-\zeta(b), 
}\\ \displaystyle{
g_2=2, g_3=3,
\wp(a)=1, \wp(b)=2, 
\wp'(a)=i, \wp'(b)=5. 
}
\end{array}
\right.
\end{eqnarray}
The canonical representation $(\wp,\wp')$ of $E(\lambda)$
\begin{eqnarray}
& &
E(\lambda)=\frac{25 + (1+2 \wp(\lambda)) \wp'(\lambda)}{2(\wp(\lambda)-2)}\ccomma
\end{eqnarray}
is optimal (triple pole at the origin),
but its shifted by $a$ (triple pole at $a$),
\begin{eqnarray}
& &
E(\lambda-a)=\frac{P_4(\wp(\lambda))+P_2(\wp(\lambda))) \wp'(\lambda)}{(\wp(\lambda)-(6+5 i)/2)(\wp(\lambda)-1)^3}\ccomma
\end{eqnarray}
requires, in order to become optimal, a translation defined by the factor
of the denominator having maximal multiplicity.

\textbf{Conflict of interest}: The author of this work declares that he has
no conflicts of interest.

\section*{Funding}.

This work was partially funded by the
RGC (Research grants council), grant 106220136,
during the author's visit to HKU in 2023-2024.



\begin{thebibliography}{99}%

\bibitem{Picard1889} \'E.~Picard,              
M\'emoire sur la th\'eorie des fonctions alg\'ebriques de deux variables,
J.~math.~pures appl.~{\bf 5} (1889) 135--319.
http://gallica.bnf.fr/ark:/12148/cb343487840/date.1889
\verb+http://sites.mathdoc.fr/JMPA/PDF/JMPA_1889_4_5_A7_0.pdf+

\bibitem{Hitchin1995-Poncelet} N.J.~Hitchin,                     
Poncelet polygons and the Painlev\'e equations,
151--185,
\textit{Geometry and analysis},
ed.~Ramanan (Oxford university press, Oxford, 1995).
ISBN            0195637402
ISBN-13      9780195637403

\bibitem{DubLNM} B.~Dubrovin,                                
Geometry of 2D topological field theories,
Lecture notes in mathematics {\bf 1620} (1996) 120--348.
https://doi.org/10.1007/BFb0094793
https://arXiv.org/abs/hep-th/9407018

\bibitem{DM2000} B.~Dubrovin and M.~Mazzocco,                    
Monodromy of certain Painlev\'e-VI transcendents and reflection groups,
Invent.~math.~{\bf 141} (2000) 55--147. 
https://doi.org/10.1007/PL00005790
https://doi.org/10.1007/s002220000065
https://arXiv.org/abs/math.AG/9806056

\bibitem{Kitaev2005Dessins} A.V.~Kitaev,                                    
Dessins d'enfants, their deformations and algebraic the sixth Painlev\'e 
and Gauss hypergeometric functions, 
Algebra i Analiz 17:1 (2005) 224--275.  
St.~Petersburg Math.~J.~{\bf 17:1} (2006) 169--206.
https://doi.org/10.1090/S1061-0022-06-00899-5
http://arXiv.org/abs/nlin.SI/0309078v3

\bibitem{Kitaev2006Angers} A.V.~Kitaev,    
Remarks towards the classification of $\hbox{RS}_4^2(3)$-transformations 
and algebraic solutions of the sixth Painlev\'e equation,
199--227,
{\it Th\'eories asymptotiques et \'equations de Painlev\'e},
eds.~E.~Delabaere and M.~Loday,
S\'eminaires et congr\`es {\bf 14}
(Soci\'et\'e math\'ematique de France, Paris, 2006).
ISBN	9782856292297
http://arXiv.org/abs/math.CA/0503082

\bibitem{Andreev-Kitaev-PVI-2002} F.V.~Andreev and A.V.~Kitaev,  
Transformations ${RS}_4^2(3)$ of the ranks $\leq4$ and
algebraic solutions of the sixth Painlev\'e equation,
Comm.~math.~phys.~{\bf 228} (2002) 151--176.
https://doi.org//10.1007/s002200200653
http://arXiv.org/abs/nlin.SI/0107074 
                                   
\bibitem{Boalch-Icosa} Philip Boalch,                  
The fifty-two icosahedral solutions to Painlev\'e VI,
Journal f\"ur die reine und angewandte Mathematik {\bf 596} (2006) 183--214.
http://dx.doi.org/10.1515/CRELLE.2006.059
https://arXiv.org/abs/math/0406281v7  

\bibitem{Boalch2007Bolibrukh} P.~Boalch,       
Some explicit solutions to the Riemann-Hilbert problem,
pp.~85--112,
\textit{Differential equations and quantum groups.
Andrey A.~Bolibrukh memorial volume},
IRMA Lectures in mathematics and theoretical physics {\bf 9}, 
(European mathematical society, Z\"urich, 2007).
https://doi.org/10.4171/020-1/6           
https://arXiv.org/abs/math/0501464v2 ``other formats'' etc

\bibitem{Boalch2005Klein} P.~Boalch,       
{}From Klein to Painlev\'e via Fourier, Laplace and Jimbo,
Proc.~London Math.~Soc.~{\bf 90} (2005) 167--208.
https://doi.org/10.1112/S0024611504015011
http://arXiv.org/abs/math.AG/0308221

\bibitem{Boalch2007Highergenus} P.P.~Boalch,            
Higher genus icosahedral Painlev\'e curves,
Funk.~Ekvac.~{\bf 50} (2007) 19--32.
\hfill\break\noindent
\verb+http://fe.math.kobe-u.ac.jp/FE/FullPapers/50-1/50_19.pdf+
\hfill\break\noindent
https://arXiv.org/abs/math/0506407v2 

\bibitem{LT2014} Oleg Lisovyy and Yuriy Tykhyy,  
Algebraic solutions of the sixth Painlev\'e equation,
J.~geom.~phys.~{\bf 85} (2014) 124-–163.
https://doi.org/10.1016/j.geomphys.2014.05.010
https://arXiv.org/abs/0809.4873v2 

\bibitem{FuchsP6} R.~Fuchs,                               
Sur quelques \'equations diff\'erentielles lin\'eaires du second ordre,
\CRAS\ {\bf 141} (1905) 555--558. 
http://refhub.elsevier.com/S1631-073X(14)00173-3/bib46756368735036s1

\bibitem{PaiCRAS1906} P.~Painlev\'e,                 
Sur les \'equations diff\'erentielles du second ordre \`a points critiques fixes,
\CRAS\ {\bf 143} (1906) 1111--1117. 
\hfill\break\noindent
https://gallica.bnf.fr/ark:/12148/cb343481087/date1906

\bibitem{CMBook2} R.~Conte and M.~Musette,
{\it The Painlev\'e handbook},
Mathematical physics studies,
xxxi+389 pages (Springer Nature, Switzerland, 2020).
https://doi.org/10.1007/978-3-030-53340-3 

\bibitem{PCI} Michel Planat, David Chester and Klee Irwin, 
Dynamics of Fricke-Painlev\'e VI surfaces,
Dynamics {\bf 4} (2024) 1--13. 
https://doi.org/10.3390/dynamics4010001

\bibitem{Kitaev1991} A.V.~Kitaev,                     
Quadratic transformations for the sixth Painlev\'e equation,
Lett.~Math.~Phys.~{\bf 21} (1991) 105--111.
https://doi.org/10.1007/BF00401643

\bibitem{TOS} T.~Tsuda, K.~Okamoto and H.~Sakai,      
Folding transformations of the Painlev\'e equations,
Math.~Annalen {\bf 331} (2005) 713--738.
https://doi.org/10.1007/s00208-004-0600-8

\bibitem{Boalch2010} Philip Boalch,  
Towards a non‐linear Schwarz's list,
in \textit{The many facets of geometry: A tribute to Nigel Hitchin},
eds.~Oscar Garcia-Prada, Jean-Pierre Bourguignon and Simon Salamon
(Oxford Scholarship Online, September 2010).
https://doi.org/10.1093/acprof:oso/9780199534920.003.0011
https://arXiv.org/abs/0707.3375v2 

\bibitem{GLBook} V.I.~Gromak and N.A.~Lukashevich,                   
{\it The analytic solutions of the Painlev\'e equations},
157 pages (in Russian),
(Universitetskoye Publishers, Minsk, 1990).
ISBN 5--7855--0319--0.

\bibitem{Okamoto1986I} K.~Okamoto,             
Studies on the Painlev\'{e} equations,
I, Sixth Painlev\'e equation, Ann.~Mat.~Pura Appl.~{\bf 146} (1986) 337--381.
https://doi.org/10.1007/BF01762370

\bibitem{CM2001c} R.~Conte and M.~Musette,
First degree birational transformations of the Painlev\'e equations
and their contiguity relations,
J.~Phys.~A {\bf 34} (2001) 10507--10522. 
http://dx.doi.org/10.1088/0305-4470/34/48/315
http://arXiv.org/abs/nlin.SI/0110028

\bibitem{Manin1998} Yu.I.~Manin,                    
Sixth Painlev\'e equation, universal elliptic curve and mirror of $P^2$,
\textit{Geometry of differential equations},
eds.~A.~Khovanskii, A.~Varchenko and V.~Vassiliev,
AMS Transl., ser. 2, {\bf 186 (39)} (1998) 131--151.
http://dx.doi.org/10.1090/trans2/186 | 
ISBN : 0-8218-1094-4   0821810944
Providence, R.I. : American Mathematical Society. 
http://arXiv.org/abs/Alg-geom/9605010 

\bibitem{Hermite-sum-zeta} Charles Hermite,  
Remarques sur la d\'ecomposition en \'el\'ements simples des fonctions
doublement p\'eriodiques,
Annales de la facult\'e des sciences de Toulouse {\bf II} (1888) C1--C12.
{\it O$\!$euvres d'Hermite}, vol IV, pp 262--273.
\verb+http://www.numdam.org/item/AFST_1888_1_2__C1_0/+

\end{thebibliography}
\end{document}